\pdfoutput=1
\documentclass{amsart}
\usepackage{amssymb,graphicx,siunitx,enumerate,url,booktabs,subfigure}
\usepackage[usenames]{xcolor}

\usepackage[normalem]{ulem}

\newcommand{\1}{{\textrm{1} \kern -.41em \textrm{1} }}

\newcommand{\ointerval}[1]{\ensuremath{\left]#1\right[}}
\newcommand{\vect}[1]{\boldsymbol{#1}}

\begin{document}
\title[Generalized balanced power diagrams for polycrystals]{Generalized balanced power diagrams for 3D representations of polycrystals\footnote{{NOTICE: this is the authors version of a work that was accepted for publication in \textbf{Philosophical Magazine}. Changes resulting from the publishing process, such as peer review, editing, corrections, structural formatting, and other quality control mechanisms may not be reflected in this document. Changes may have been made to this work since it was submitted for publication. A definitive version appears in Philosophical Magazine.}}}

%

\author[A. Alpers]{Andreas Alpers}
\address[Andreas Alpers]{Zentrum Mathematik, Technische Universit\"at M\"unchen, D-85747 Garching bei M\"unchen, Germany}
\email{alpers@ma.tum.de}
\thanks{Corresponding author: A. Alpers}
\author[A. Brieden]{Andreas Brieden}
\address[Andreas Brieden]{Fakult\"at f\"ur Wirtschafts- und Organisationswissenschaften, Universit\"at der Bundeswehr M\"unchen, D-85579 Neubiberg, Germany}
\email{andreas.brieden@unibw.de}
\author[P. Gritzmann]{Peter Gritzmann}
\address[Peter Gritzmann]{Zentrum Mathematik, Technische Universit\"at M\"unchen, D-85747 Garching bei M\"unchen, Germany}
\email{gritzman@ma.tum.de}
\author[A. Lyckegaard]{Allan Lyckegaard}
\address[Allan Lyckegaard]{Xnovo Technology, Galoche All{\'e} 15, Dk- 4600 Koege, Denmark}
\email{alyckegaard@xnovotech.com}
\author[H.F. Poulsen]{Henning Friis Poulsen}
\address[Henning Friis Poulsen]{NEXMAP, Department of Physics, Technical University of Denmark, 2800 Lyngby, Denmark}
\email{hfpo@fysik.dtu.dk}


\maketitle

\begin{abstract} 
Characterizing the grain structure of polycrystalline material is an important task in material science. 
The present paper introduces the concept of generalized balanced power diagrams as a concise alternative to voxelated mappings.
Here, each grain is represented by (measured approximations of) its center-of-mass position, its volume and, if available, by its second-order moments
(in the non-equiaxed case). Such parameters may be obtained from 3D x-ray diffraction. As the exact global optimum of our model results from the solution of a suitable linear program it can be computed quite efficiently. Based on verified real-world measurements we show that from the few parameters per grain (3, respectively 6 in 2D and 4, respectively 10 in 3D) we obtain excellent representations of both equiaxed and non-equiaxed structures. Hence our approach seems to capture the physical principles governing the forming of such polycrystals in the underlying process quite well.
\end{abstract}


\section{Introduction}
During the last decade, the field of 3D materials science has emerged and matured~\cite{thortonpoulsen08}. A prime objective is to relate experimental 3D grain maps with 3D simulations, e.g. in the field of phase transformations, plasticity or grain growth.  Coupling serial sectioning strategies with electron microscopy, 3D maps of grain orientations can be acquired with a resolution as good as \SI{20}{\nano\metre} \cite{zaefferer08, uchic06}.
For nanoncrystalline materials, 3D grain mapping of thin foils at the \SI{1}{\nano\metre} scale has been reported using a transmission electron microscope (TEM)~\cite{liu11}. On the other hand, synchrotron based x-ray diffraction methods such as 3-dimensional x-ray diffraction (3DXRD)~\cite{poulsennielsen01, hefferan09, additional1}, diffraction contrast tomography (DCT)~\cite{ludwig08, king08, additional2}, differential-aperture x-ray microscopy (DAXM)~\cite{larson02, levine06, ice11} and scanning diffraction tomography~\cite{bleuet08} are non-destructive and allow for an acquisition of 3D movies of grain and sub-grain evolution.  Using a map of the initial state of the sample as input to 3D simulations, one may compare the evolution in the simulated movie to the experimental one point-by-point and time-step by time-step~\cite{mckenna14}. This is seen as a new and powerful route to elucidate the underlying materials physics and to verify materials models. 

The conventional representation of such 3D grain maps is by their individual voxels; see \cite{dream3d1, dream3d2} for a simulation tool. This representation is heavy in terms of computing and, more importantly, often out of reach experimentally.  In particular for the time resolved x-ray studies a compromise is often required between time and spatial resolution. Frequently, the exact grain shape is not measured, e.g. due to asymmetric (3DXRD) or poor spatial resolution (use of far-field detectors) or use of mapping strategies involving integration over more than one spatial dimension---so-called boxscans \cite{lyckegard10, phdallan}. In these cases what can be measured with high fidelity is typically the center-of-mass (CMS) position and the volume of each grain as well as information about grain breadths in some directions. In the following we will focus on information about second-order moments, which are readily available, e.g., through boxscan methods taken from several directions as the images contain the elongation in the correspondingly projected grain; see \cite[Section~3]{phdallan}. Our method is, however, flexible in terms of using other kinds of information. 

This situation motivates a revisit of the suitability of tessellations as a way of representing grain maps.  Previous studies have focused on heuristics for computing Laguerre tessellations, also known as power diagrams, where grains are characterized by seed points and volumes~\cite{kuehn08, laguerre11, laguerre2, telley92, aljfs-14} (see also \cite{duan2014} for the inverse task). Such tessellations have been used in connection with models of grain growth~\cite{riosglicksman07, telley96a, telley96b, xue97} and in connection with mechanical models where it is sufficient to know which grains are neighbors~\cite{mikadawson98}. Tessellations may also be relevant in connection with topological descriptions of grains~\cite{dehoff85, patterson13}. (We emphasize that in case of access to accurate voxelated data, these evidently contain more information, and typically will provide better statistics of misorientation neighbor number distributions and boundary normals. The topic at hand is how far one can go with the much condensed description of a tessellation.)

As power diagram tessellations are composed of convex cells, more general tessellations have long been considered; see \cite{akl-13} for a comprehensive treatment of Voronoi diagrams and their generalizations. In particular, the effect of the use of local metrics has been studied in \cite{ls-03} and \cite{bwy-08}. Recently (and independently of the present paper), \cite{aljfs-14} examined the effect of various types of local metrics for deriving stochastic models for microstructures. In fact, \cite{aljfs-14} consider three different models, the most relevant for the present paper being the anisotropic growth model using ellipsoidal norms.

As its main contribution, the present paper introduces a general linear programming model, based on \cite{briedengritzmann12}, for computing (globally) \emph{optimal} tessellations very efficiently. In particular, it allows to set parameters that encode arbitrary local distance measures including local metrics. Further, the derived tessellations comply with measured grain volume information up to any required accuracy. To the best of our knowledge, this cannot be guaranteed by any other standard grain map model. We describe our approach in its most elementary form with a particular focus on ellipsoidal norms to allow at least a rough comparison with the results of \cite{aljfs-14}. This special way of setting parameters will be referred to as \emph{generalized balanced power diagrams} (GBPDs). GBPDs allow to utilize CMS, volume, and second-order information in a rigorous way. Our algorithm can, of course, easily been extended using global techniques from \cite{briedengritzmann10} or local variants of \cite{bbg13}.

We will demonstrate the method's very favorable performance on experimentally determined grain structures in 2D and 3D, and for equiaxed and non-equiaxed structures. As it turns out, the quality of fit is so high that it might be seen as an indication that our mathematical minimization model may indeed capture the dominating physical principles governing the forming of such polycrystals in the underlying processes.

\section{Mathematical Background}
In the following, let $\mathbb{R}^d$ denote the $d$-dimensional Euclidean space (most relevant for our purposes is $d \in \{2,3\}$). Further, for $k \in \mathbb{N}$ let $[k]:=\{1,\dots,k\}$. For a positive definite matrix $A$ we denote by $||\cdot||_A$ the ellipsoidal norm, defined by 
\begin{equation}\label{eq:metric}
||\vect x||_A:=\sqrt{\vect x^T A \vect x}.
\end{equation}
Note that for $A$ being the identity matrix $I$ we obtain the Euclidean norm.

Our aim is to reconstruct what we call \emph{generalized balanced power diagrams}. These diagrams generalize power diagrams, which in turn generalize Voronoi diagrams  (as we discuss further below); see also \cite[Chapter~6.2]{akl-13}. 
A generalized balanced power diagram is specified by a set of distinct sites $S:=\{\vect{s_1},\dots,\vect{s_k}\}\subseteq \mathbb{R}^d$, parameters $(\sigma_1,\dots,\sigma_k)^T\in\mathbb{R}^k$, and positive definite matrices $A_1,\dots,A_k\in\mathbb{R}^{d\times d}$. The $i$-th generalized balanced power cell $P_i$ is then defined by 
\begin{equation}\label{eq:powercell}
P_i:=\{\vect{x} \in \mathbb{R}^d\::\: ||\vect{x} - \vect{s_i}||_{A_i}^2-\sigma_i \leq ||\vect{x} -\vect{s_j} ||_{A_j}^2 -\sigma_j, \: \forall j\neq i\}.
\end{equation}
The generalized balanced power diagram $P$ is the $k$-tuple $P:=(P_1,\dots,P_k)$. Let us point out that the method given will be able to find optimal $\sigma_1,\dots,\sigma_k$ that guarantee that the volumes of the cell are within prescribed ranges.

Somewhat surprising at first glance, generalized balanced power diagrams are closely related to optimal clusterings; see \cite{briedengritzmann12}. For this we introduce a particular clustering method that is based on solving a \emph{weight-balanced least-squares assignment} problem. This particular assignment problem is specified by a set of points $X:=\{\vect{x_1},\dots,\vect{x_m} \}\subseteq \mathbb{R}^d$, sites $S:=\{\vect{s_1},\dots,\vect{s_k}\}\subseteq\mathbb{R}^d$, weights $\omega_1,\dots,\omega_m \in \ointerval{0,\infty}$,  positive definite matrices $A_1,\dots,A_k\in\mathbb{R}^{d\times d}$, and cluster size bounds $\vect \kappa^-:=(\kappa_1^-,\dots,\kappa_k^-)^T$, $\vect \kappa^+:=(\kappa_1^+,\dots,\kappa_k^+)^T$ with $0< \kappa_i^-\leq  \kappa_i^+$ and \[\sum_{i=1}^k\kappa_i^-\leq\sum_{j=1}^m\omega_j\leq \sum_{i=1}^k\kappa_i^+.\] 
 The (fractional)  \emph{weight-balanced least-squares assignment} problem is the following linear optimization problem:
\begin{equation}\label{eq:lp}
\begin{array}{lll}
\textnormal{(LP)}        &\min \sum_{i=1}^k\sum_{j=1}^m\gamma_{i,j}\xi_{i,j}     &\\
\textnormal{subject to}  &\sum_{i=1}^k\xi_{i,j}=1                                 &(j\in[m]),\\
                         &\kappa_i^-\leq\sum_{j=1}^m \xi_{i,j}\omega_j \leq \kappa_i^+               &(i\in[k]),\\
                         &\xi_{i,j}\geq 0                                         &(i\in[k];\: j\in[m]),
\end{array}
\end{equation}
with $\gamma_{i,j}:=\omega_j||\vect{x_j}-\vect{s_i}||_{A_i}^2$ for all $i,j$. The $\xi_{i,j}$ are the variables; they specify the fraction of point $\vect{x_j}$ that is assigned to site $\vect{s_i}$. Any optimal solution $C:=(C_1,\dots,C_k)$ of~\eqref{eq:lp} where $C_i:=(\xi_{i,1},\dots,\xi_{i,m})$ is called a \emph{(fractional) weight-balanced least-squares assignment for $X$ with sites $\{\vect{s_1},\dots,\vect{s_k}\}$}. We remark that the parameters $\gamma_{i,j}$ are the only elements that depend on the particular norms. Hence (LP) can be easily adapted to use other distance measures and incoporate other information on the grains. We will further comment on this point at the end of the paper.

In the particular case of unit weights $(\omega_1,\dots,\omega_m)=(1,\dots,1)=:\1^T$ we have a totally unimodular constraint matrix, which implies that $0/1$-solutions can be found as basic feasible solutions (for a definition see, e.g., \cite[Def.~2.4]{papadimitriou98}) in polynomial time. We remark that voxelized maps can be obtained with this approach if the $\vect{x}_j$ represent voxels (see Section~\ref{sect:3} for further details). In fact, $\xi_{i,j}=1$ in a solution of \eqref{eq:lp} means that $\vect{x}_j$ belongs to the $i$th grain.
These voxelized maps, in fact, represent generalized balanced power diagrams.

Generalized balanced power diagrams generalize \emph{Voronoi} and \emph{power diagrams}; the latter are also known as  \emph{Laguerre} or \emph{Dirichlet tessellations}. Both Voronoi and power diagrams are convex cell decompositions where the cells are given by~\eqref{eq:powercell}. 
For power diagrams we have the restriction $A_1=\cdots=A_k$ while the classical Voronoi diagrams are obtained with $A_1=\cdots=A_k=I$ and $\sigma_1=\cdots=\sigma_k$. For arbitrary positive definite matrices, however, the generalized balanced power cell i.e. the $P_i$ of \eqref{eq:powercell} need not be convex. 

\section{Approach}\label{sect:3}
In the model introduced above, we optimize the objective function in~\eqref{eq:lp}, with fixed sites $\vect{s_1},\dots,\vect{s_k}$ (representing the measured CMS of the grains), and fixed values for the second-order moments, which define the metric for each grain, \eqref{eq:metric}. The area or volume $\kappa_i$ of each grain is allowed to vary within a certain range $\kappa_i\pm\varepsilon$. Accordingly in~\eqref{eq:lp} we set $\kappa_i^-:=\kappa_i-\varepsilon$ and $\kappa_i^+:=\kappa_i+\varepsilon$. In our implementation we choose $\varepsilon:=2$, i.e., the grain volumes (resp. areas) are allowed to vary by $\pm 2$ voxels (resp. pixels). Further, we choose $(\omega_1,\dots,\omega_m)=\1^T$ as this leads to solutions where each voxel is associated with exactly one grain. 

For our test cases, the ellipsoidal norms are derived from a principal component analysis of the grains in the following way: suppose for a set of points 
$\{\vect {g_1}, \dots, \vect {g_l}\} \subseteq \mathbb{R}^d$ of a grain we are given the principal components $\vect{u_1},\dots,\vect{u_d}$ and corresponding eigenvalues $\lambda_1,\dots,\lambda_d$ of the $d\times d$ covariance matrix of $G=(\vect{g_1},\dots,\vect{g_l})$. The norm $||\cdot||_A$, where
\begin{equation} \label{eq:amatrix}
A:=U\Lambda^{-1}U^T
\end{equation}
with $U:=(\vect{u_1},\dots,\vect{u_d})$ orthogonal and $\Lambda:=\textnormal{diag}(\lambda_1,\dots,\lambda_d)$ is an ellipsoidal norm for which $\{\vect{x}\in\mathbb{R}^d\::\: ||\vect{x}||_A=1 \}$ defines an ellipsoid with semi-axes $\vect{u_i}$ of lengths $\sqrt{\lambda_i}$, $i=1,\dots,d$. 

It is worth noting that our approach generates by design (generalized balanced) power cells whose volumes (resp. areas) lie in the prescribed range. The centers of the cells, however, are not automatically guaranteed to coincide with $\vect{s_1},\dots,\vect{s_k}$. The implications of this is discussed below.   

\section{Samples}\label{sect:samples}
We consider two data sets, in the following referred to as Data Set~I and Data Set~II. Data Set~I exemplifies the ``isotropic'' case which can be represented very accurately already by the centers and volumes (resp. areas) of the grains. (In the formulation \eqref{eq:lp} this means, of course, that
$A_1=\cdots=A_k=I$.) The ``anisotropic'' Data Set~II, however, requires moment information and makes use of the enriched concept of generalized balanced power diagrams as here the matrices $A_i$ of the individual grains are quite different. The following two paragraphs give more details. 

Data Set~I is taken from \cite{laguerre11} and represents a real equiaxed 3D grain structure obtained by a synchrotron micro-tomography experiment conducted on a meta-stable beta titanium alloy (Ti~$\beta$21S). 
This data set has been used for validation of 3DXRD and DCT data since the grain boundaries were decorated by precipitation of a second phase that allowed a direct 3D imaging of the grain shapes using propagation based phase contrast tomography \cite{validationCT}. For the data set a \SI{300}{\micro\metre} cylinder-shaped sample was scanned with a resolution resulting in a voxel size of \SI{0.7}{\micro\metre} in the final 3D reconstruction (we refer to this resolution as \emph{full resolution}). A subvolume of size 
\SI{240 x 240 x 420}{\micro\metre} was extracted from the reconstruction for analysis resulting in a $339\times339\times599$ voxel volume. The subvolume has $591$ grains with $211$ interior grains. 
The ratio between the average lengths of the largest and smallest semiaxes of the ellipsoids obtained from a principal component analysis of the grains is $1.6$; the grains are rather equiaxed. The grains have an average width of \num[group-separator = {,}]{73.6} voxels in the longest and \num[group-separator = {,}]{47.6}  voxels in the smallest semiaxis direction; on average, interior grains consist of \num[group-separator = {,}]{130156.8} voxels. 

Data Set~II represents a real non-equiaxed 2D grain structure measured by electron backscatter diffraction (EBSD) in a thick plate of commercial purity aluminum (AA 1050) heavily deformed by 8 passes of equal channel angular extrusion (ECAE) with sequential \SI{90}{\degree} rotations about the plate normal direction~\cite{oleg2012}. The EBSD map used in the present work has a size of \SI{8.5 x 8.5}{\micro\metre} and we aim at a final 2D reconstruction with a pixel resolution of \SI{25}{\nano\metre} resulting in a $339 \times 339$ pixel image. The EBSD map contains 206 grains (they are in fact subgrains of an initial \SI{50}{\micro\metre} sized grain) with 140 interior grains. The ratio between the average lengths of the two semiaxes of the ellipsoids obtained from a principal component analysis of the grains is $2.93$. The grains in the $339 \times 339$ discretization have an average width of $43.3$ pixels in the longer and $16.2$ pixels in the smaller semiaxis direction; on average, interior grains consist of $600$ pixels. For the power diagrams we set $A_1=\cdots=A_k$ with $A_1$ obtained as (area-) weighted average of the individual $A$ matrices as in~\eqref{eq:amatrix}; the resulting matrix $A_1$ has eigenvalues $0.0546$ and $0.0119$.

For both data sets we computed representations based on the known heuristic (H) of \cite{laguerre11} that produces a power diagram, on optimal power diagrams (PD) according to~\eqref{eq:lp}, and on generalized balanced power diagrams (GBPD). For Data Set~I, PD was produced without any moment information i.e., we simply used the Euclidean distance while for Data Set~II a uniform matrix $A$ was employed for each grain. Hence for PD \eqref{eq:lp} produces an optimal power diagram for both data sets where each grain is represented by a convex polyhedral cell. 

\section{Results}\label{sect:4}
In this section we present 2D, lower resolution, and 3D tessellations of the data sets. We focus on the power diagram and the general approach. Approximations for the power diagrams obtained by the heuristic from \cite{laguerre11} are included for comparison. All axes labels of the presented images are given in voxel (resp. pixel) units. We give several statistical parameters to compare the quality of different tessellations. Repeatedly, we shall present the parameters in the form $E=(e_1,\dots,e_6)$, with $e_1$ denoting the percentage of correctly labeled voxels, $e_2$ denoting the percentage of grains with all neighbors correctly reconstructed, $e_3$ giving the percentage of grains that have at most one incorrectly reconstructed neighbor, $e_4$ representing the number of erroneous extra neighbors per grain, $e_5$ giving the number of erroneous missing neighbors per grain, and with $e_6$ denoting the total number of erroneous neighbors per grain. The computations were performed on a standard laptop (Intel Core i5-2520M 2.3 GHz with 4 GByte RAM). As a linear programming solver we used Xpress version $7.5$; see \cite{Xpress}.

\subsection{2D tessellations}
Table~\ref{table1} and~\ref{table2} give several statistical parameters for the 2D tessellation results. Table~\ref{table1} focuses on the general reconstruction quality and grain topology (the data columns present the above mentioned parameters $E=(e_1,\dots,e_6)$), while Table~\ref{table2} lists the average error (and standard deviation) of the reconstructed areas and the Euclidean distance between CMS and the reconstructed CMS. 

In all cases we see an improvement in the statistical parameters while we move from the heuristic (H), over the power diagram (PD), to the generalized balanced power diagram (GBPD). The largest improvement of GBPD over H and PD can be found for Data Set~II (non-equiaxed grains). For Data Set~I we remark that the average CMS displacement in the PD and GBPD approach at full resolution is not larger than \SI{2.3}{\micro\metre}; this is of the same order as the average measurement error on the CMS in the data (which for typical grains in this sample is about \SI{2.6}{\micro\metre}). The average errors of the reconstructed grain areas in the PD and GBPD approach are very small as is enforced by the constraints in~\eqref{eq:lp}. 

\begin{table}[htb]
\caption{Statistical parameters for a comparison of the tessellation results. The H, PD, GBPD columns refer to the heuristic from~\cite{laguerre11}, the power diagram approach ($A_1=\cdots=A_k$), and the generalized balanced power diagram approach, respectively.}\label{table1}
\centering
\begin{tabular}{@{} *1l  *3r  p{1ex} *3r }    \toprule
 & \multicolumn{3}{c}{Data Set~I} && \multicolumn{3}{c}{Data Set~II} \\[0.7ex]
& \multicolumn{1}{c}{H} & \multicolumn{1}{c}{PD} & \multicolumn{1}{c}{GBPD}  && \multicolumn{1}{c}{H} & \multicolumn{1}{c}{PD} & \multicolumn{1}{c}{GBPD}   \\\midrule
 \% Correctly labeled voxels               & 90.51  & 91.08 & 96.12 && 61.96  & 76.71 & 93.89\\ 
 \% Grains with all neighbors correct    & 61.33  & 62.52  & 74.04 && 3.40  & 39.32 & 79.61\\ 
 \% Grains with $\leq 1$ incorrect neighbor   & 90.40  & 92.44  & 96.34 && 16.02  & 70.87 & 96.12\\  
 \# Erroneous extra neighbors/grain    & 0.21   & 0.20 & 0.11 && 1.47  & 0.60 & 0.12\\
 \# Erroneous missing neighbors/grain  & 0.30   & 0.27  & 0.19 && 2.17  & 0.56 & 0.13\\ 
 \# Erroneous neighbors/grain       & 0.51  & 0.47  & 0.30 && 3.64  & 1.16 & 0.25\\ \bottomrule
 \hline\\
\end{tabular}
\end{table}

\begin{table}[htb]
\caption{The average area error and the average CMS displacement (i.e., average Euclidean distance between CMS and reconstructed CMS). Standard deviations are given in parentheses. All values are given in pixel units. The H, PD, GBPD columns refer to the heuristic from~\cite{laguerre11}, the power diagram approach, and the generalized balanced power diagram approach, respectively.}
\label{table2}
\centering
\begin{tabular}{@{} *1l  *3r   }    \toprule
& \multicolumn{1}{c}{H} & \multicolumn{1}{c}{PD} & \multicolumn{1}{c}{GBPD}     \\\midrule
Data Set I  &&&\\
  \hspace*{2ex}Average area error                 &178.77 (155.39) & 1.98 (0.15) & 1.92 (0.33)\\ 
 \hspace*{2ex}Average CMS displacement      & 4.70\hspace*{2ex}(19.70)   &3.28 (2.51) & 0.68 (0.66)\\ 
Data Set II  &&&\\
  \hspace*{2ex}Average area error                 &152.69 (159.65) & 1.98 (0.14) & 1.81 (0.53)\\  
  \hspace*{2ex}Average CMS displacement    & 35.70 \hspace*{1ex}(91.45)   & 4.16 (3.91) & 0.50 (0.53)\\  
   \bottomrule
 \hline\\
\end{tabular}
\end{table}

\begin{figure}[htb]
\subfigure[]{\includegraphics[height=0.3\textheight]{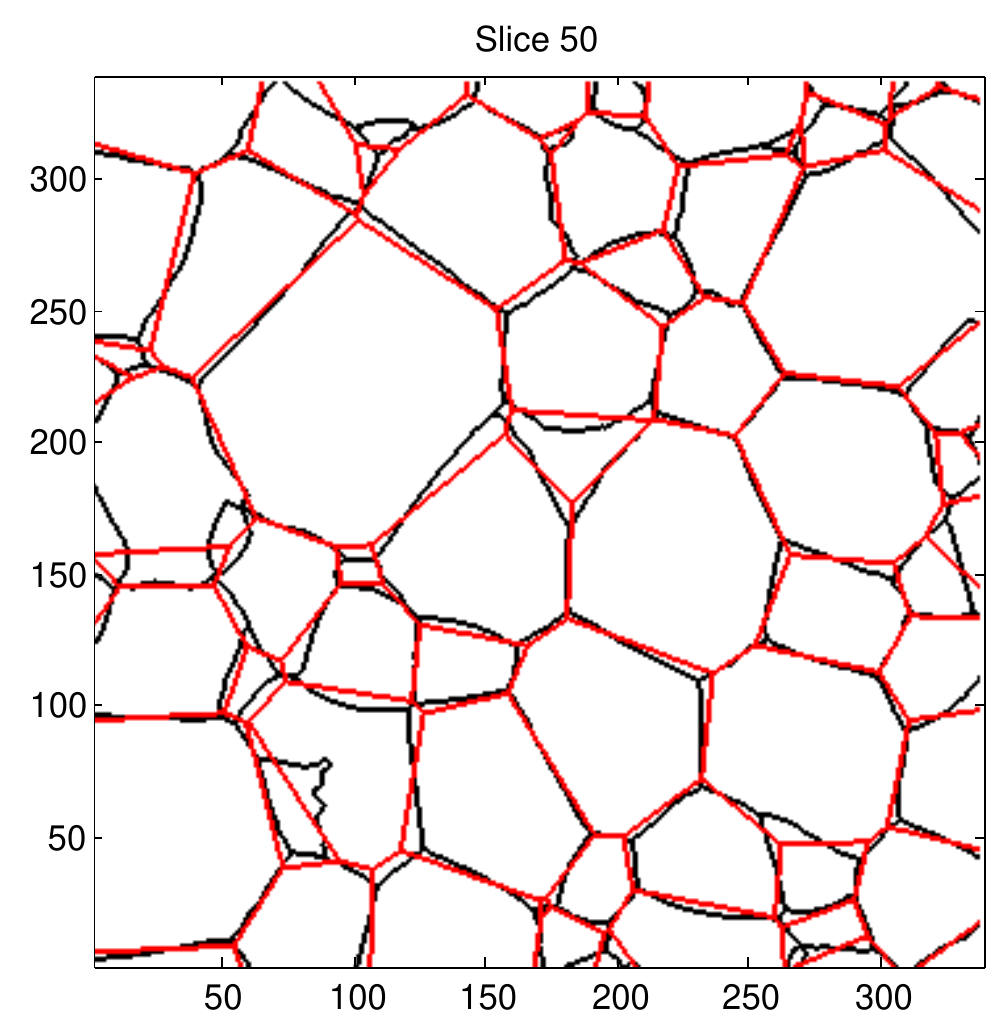}}
\subfigure[]{\includegraphics[height=0.3\textheight]{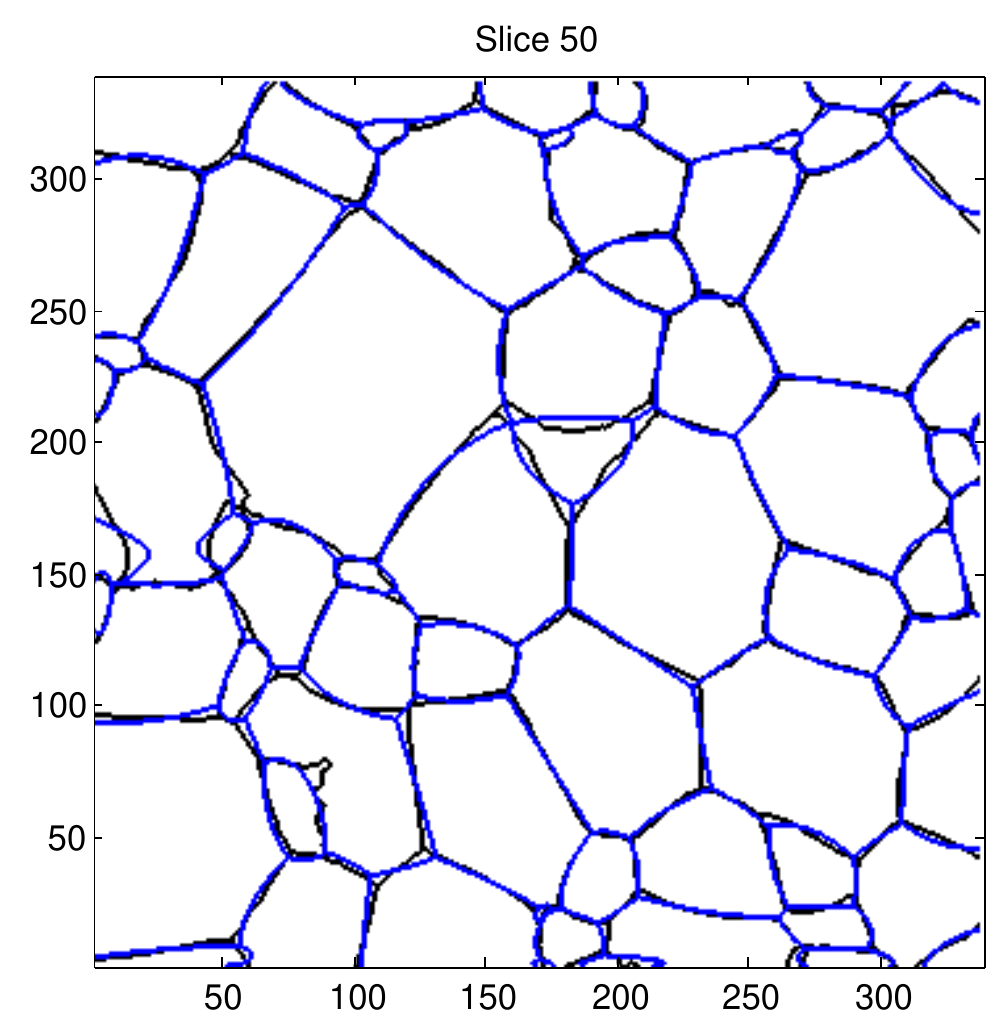}}
\caption{Tessellations for a typical slice (here, precisely slice~50) that contains several non-convex grains. Real grain boundaries are shown in black, (a)~the power diagram reconstruction is given in red, (b)~the generalized balanced power diagram reconstruction is shown in blue.}\label{fig4a}
\end{figure}

\begin{figure}[htb]
\subfigure[]{\includegraphics[height=0.3\textheight]{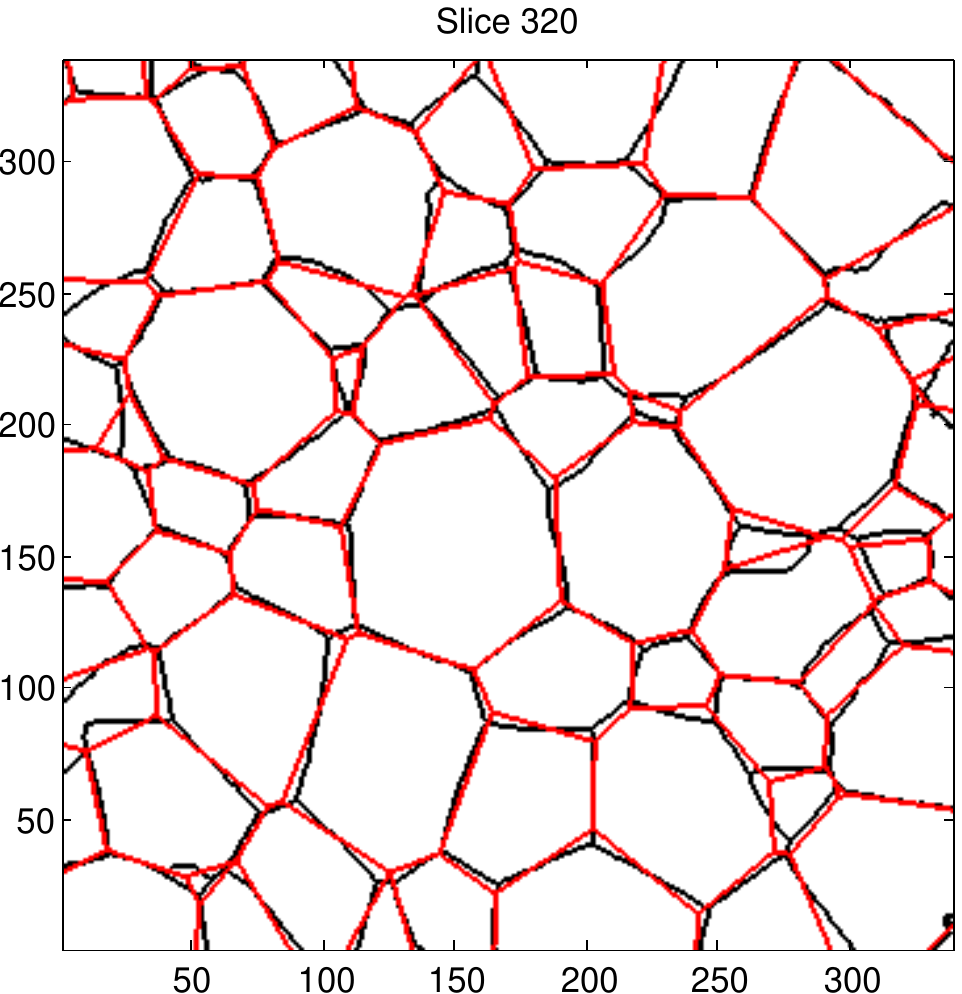}}
\subfigure[]{\includegraphics[height=0.3\textheight]{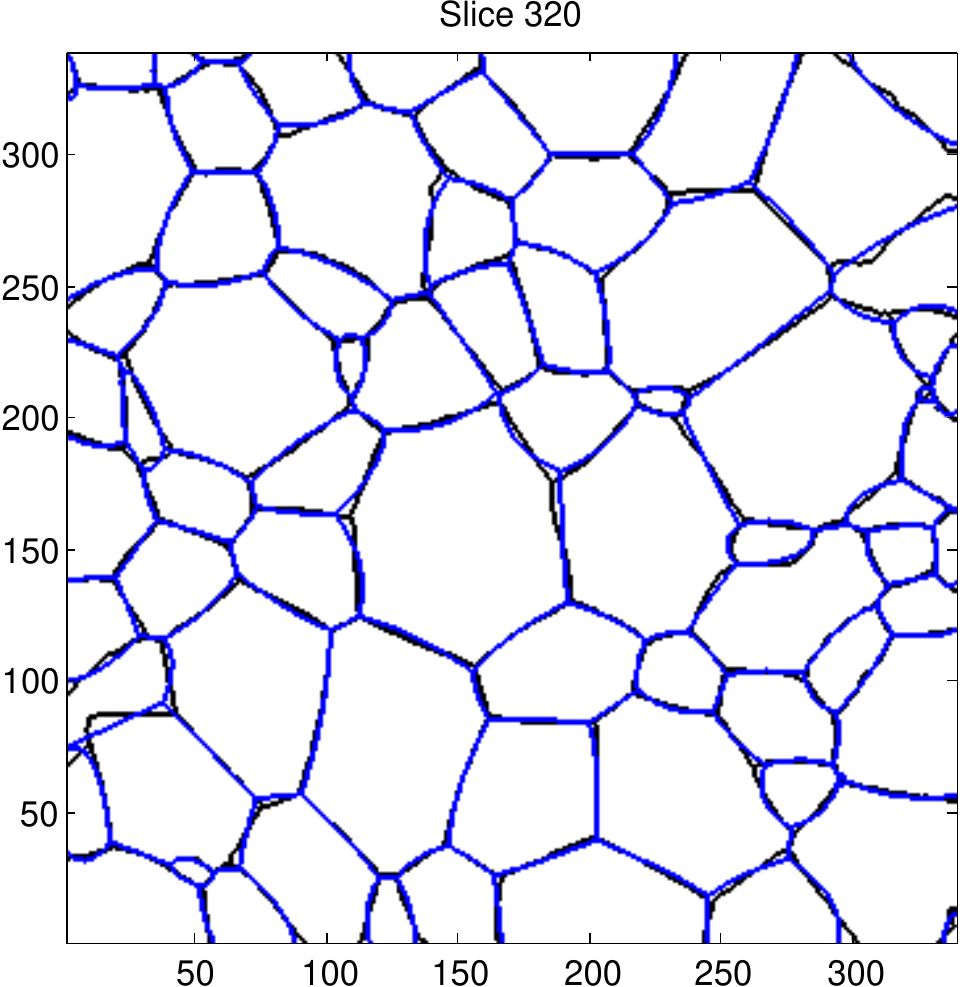}}
\caption{Tessellations for a typical slice (here, precisely slice~320) that contains mainly convex grains. Real grain boundaries are shown in black, (a)~the power diagram reconstruction is given in red, (b)~the generalized balanced power diagram reconstruction is shown in blue.}\label{fig5a}
\end{figure}

\begin{figure}[htb]
\begin{center}
\subfigure[]{\includegraphics[height=0.2\textheight]{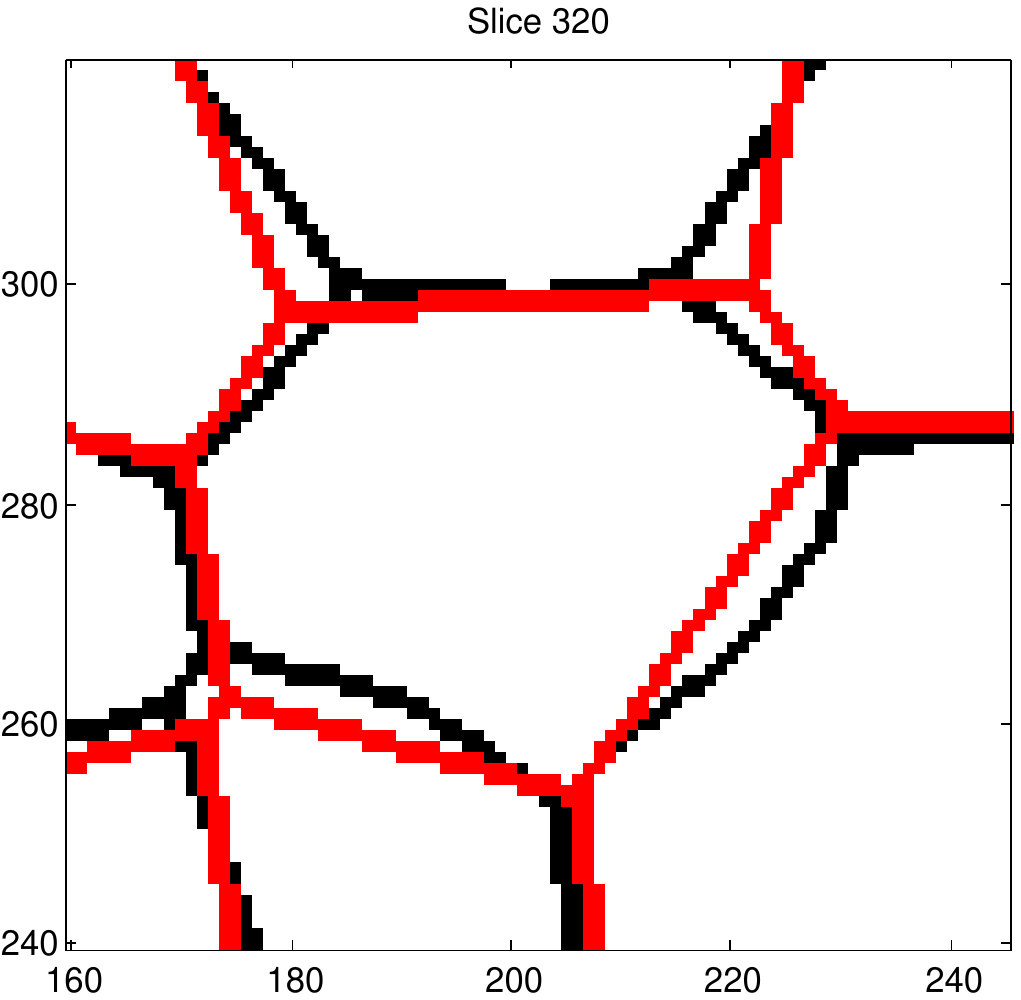}}
\subfigure[]{\includegraphics[height=0.2\textheight]{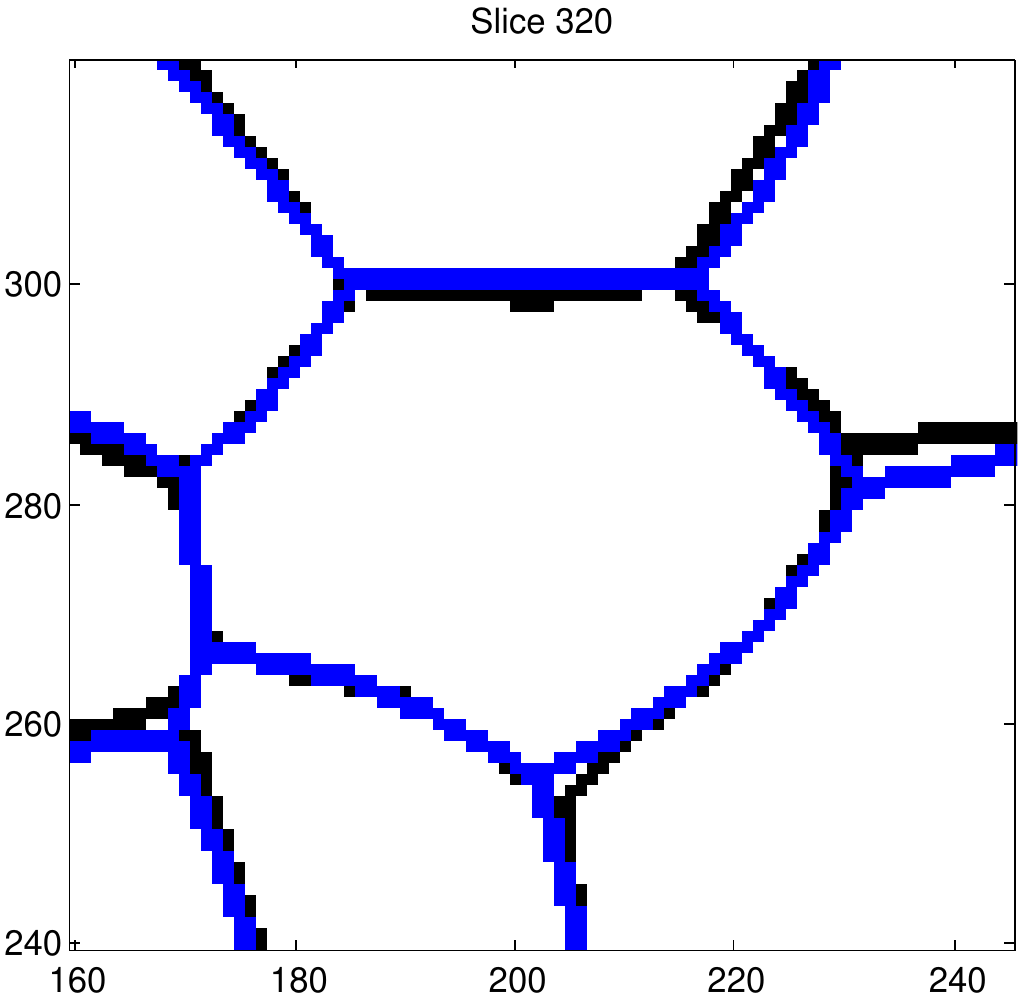}}
\end{center}
\caption{Magnifications of central parts of Fig.~\ref{fig5a}.}\label{fig5b}
\end{figure}

\begin{figure}[htb]
\subfigure[]{\includegraphics[height=0.3\textheight]{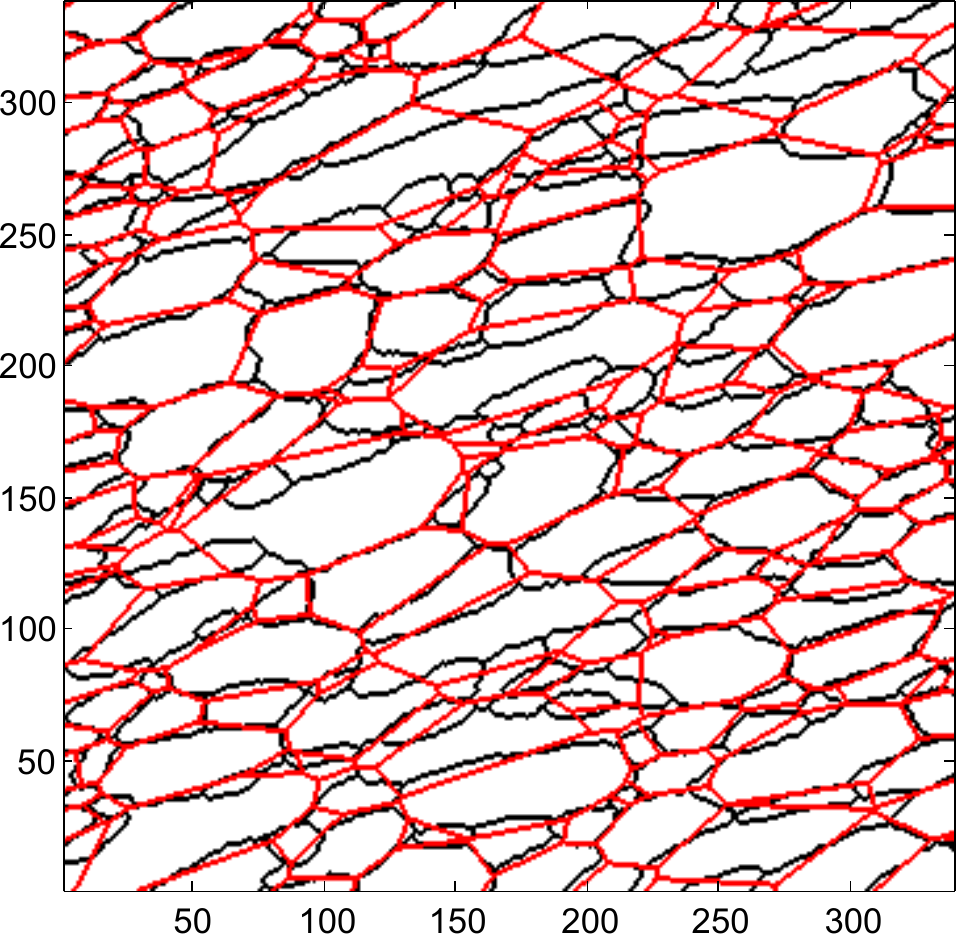}}
\subfigure[]{\includegraphics[height=0.3\textheight]{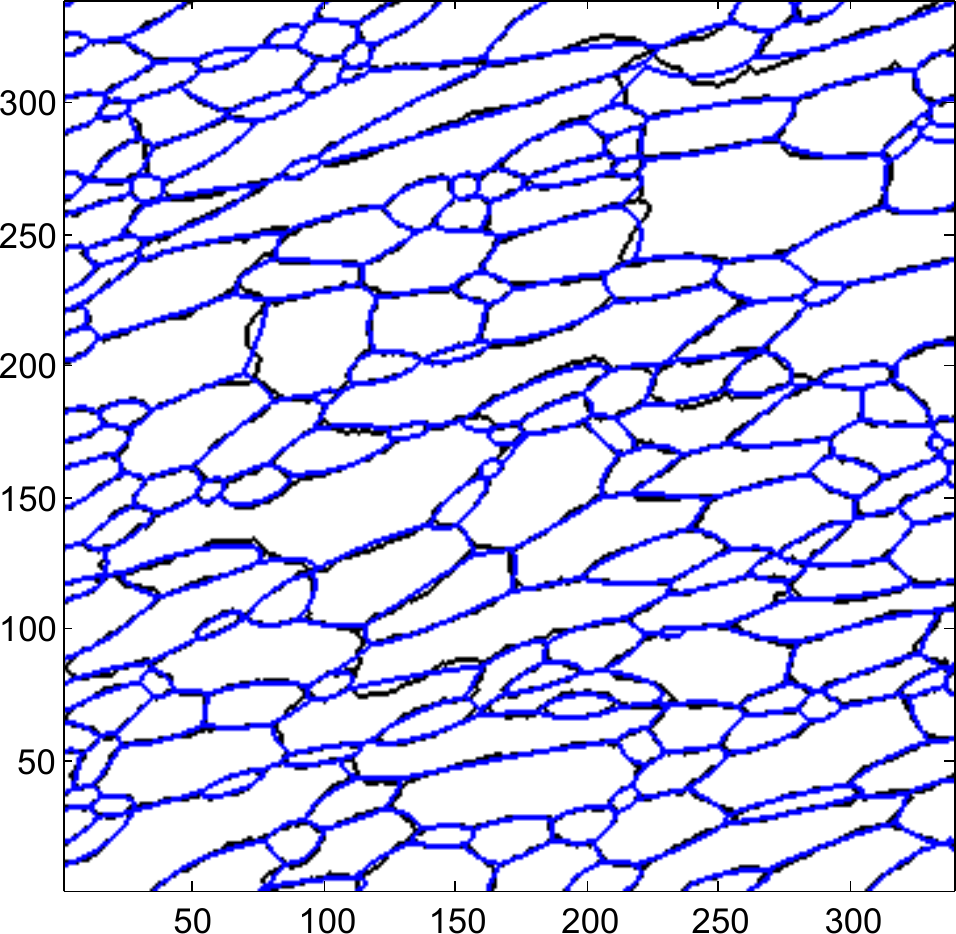}}
\caption{Tessellations for Data Set~II that contains many non-equiaxed grains. Real grain boundaries are shown in black, (a)~the power diagram reconstruction is given in red, (b)~the generalized balanced power diagram reconstruction is shown in blue.}\label{fig8}
\end{figure}

In Figs.~\ref{fig4a}-\ref{fig8} we show tessellations for several particular slices at full resolution. 
Note that the maps of the $k$ grains are characterized by a total of $3k$ and $3k+3$ parameters for PD in Data Set~I and II, respectively, whereas
GBPD uses $6k$ parameters. Since PD produces convex polyhedral cells, it should not come as a surprise that in the anisotropic case of Data Set~II
the largest deviations from the real grain map appear near non-convex grain boundaries.

As the numbers of used parameters are small, the extremely good fit for GBPD on Data Set~II and even of PD on Data Set~I suggests that
our model does indeed capture the underlying physics of the forming process of polycrystals. This, however, requires further investigation.

In an additional numerical experiment, we reconstructed slices of Data Set~II by varying the $\varepsilon$ in the cluster size bounds $\kappa_i^-:=\kappa_i-\varepsilon$ and $\kappa_i^+=\kappa_i+\varepsilon$. In particular, we performed $10$ GBPD reconstructions with $\varepsilon:=\varepsilon(t):=2+10t$ in the $t$-th reconstruction. While for $\varepsilon(1)$ we have $E=(93.72, 76.21, 96.60, 0.12, 0.16, 0.28)$, we obtain for increasing $\varepsilon$ only slightly poorer reconstructions; the result for $\varepsilon(10)$, for instance, is $E=(93.34, 76.70, 96.12, 0.13, 0.15, 0.28)$. This indicates a rather stable behavior of the reconstruction problem for this data set under uniform changes of the cluster size bounds. It is expected that the cluster size bounds have a larger effect for irregular structures.

Clearly, the possibility of a comparison of the results given above with those of \cite{aljfs-14} is rather limited. However, it may be fair to say that the deviations found on our data set with our method are significantly better than those reported by \cite{aljfs-14} for their data sets.

\subsection{Lower resolution tessellations}
To test the stability of the obtained solutions and the possibility of speed-ups we performed independent reconstructions with PD and GBPD at half resolution (\SI{1.4}{\micro\metre} voxel size for Data Set~I and \SI{50}{\nano\metre} pixel size for Data Set~II). 

 While a typical problem instance for a full resolution reconstruction of a slice for Data Set~I with $60$ grains contains \num[group-separator = {,}]{6895260} variables and \num[group-separator = {,}]{7010301} constrains (including the non-negativity constraints), the half resolution reconstruction in 2D involves only $1/4$ of the full resolution variables. Running times for typical problem instances at full resolution amount to \SIrange[range-phrase=--,range-units=single]{5}{7}{\minute}, while running times for half resolution instances amount to about \SI{30}{\second}.

\begin{figure}[htb]
\subfigure[]{\includegraphics[height=0.3\textheight]{slice320_generalpower.pdf}}
\subfigure[]{\includegraphics[height=0.3\textheight]{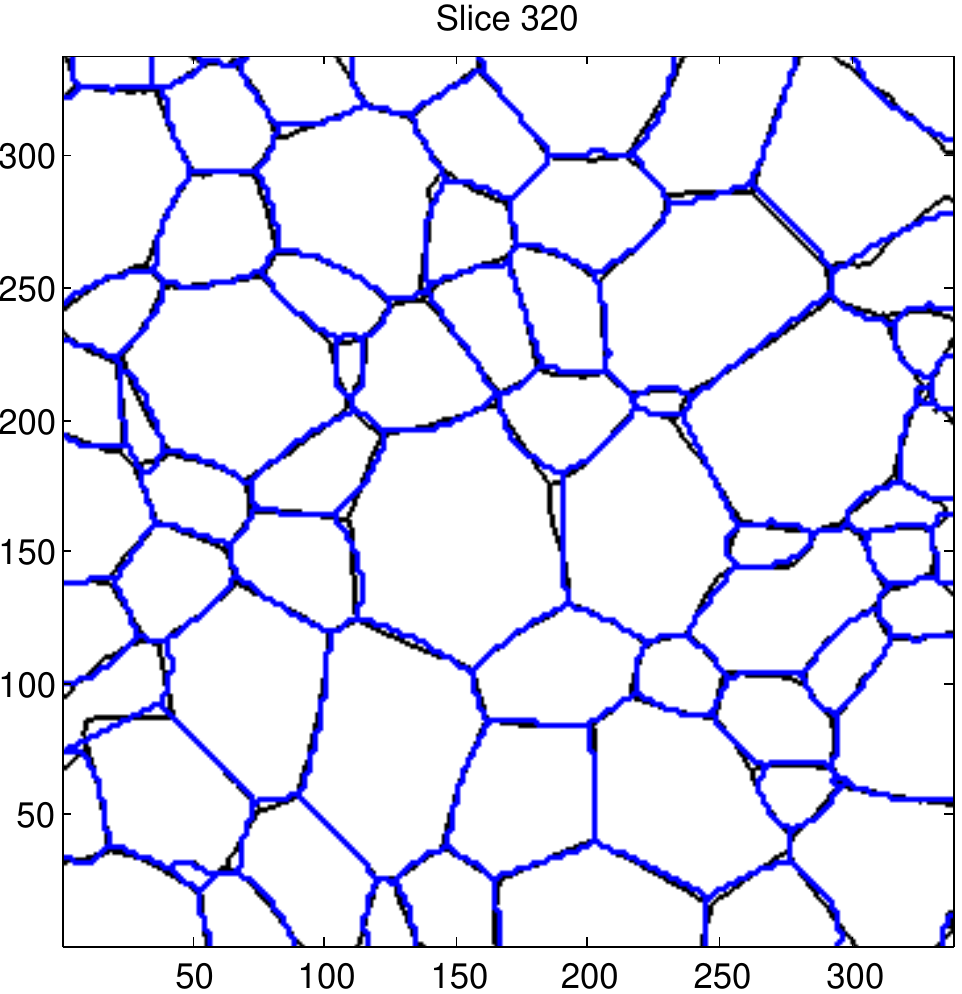}}
\caption{Generalized balanced power diagrams for slice 320 obtained at (a)~full resolution and (b)~half resolution. The real grain boundaries at full resolution are depicted in black color.}\label{fig7}
\end{figure}

The reconstructions obtained at half resolution seem not to deviate much from the reconstructions at full resolution. Fig.~\ref{fig7} shows a typical slice and reconstruction for Data Set~I with GBPD at full and half resolution. Results for Data Set~II and, respectively, PD are similar. For PD on Data Set~I with half resolution we obtain $E=(90.69, 60.86, 91.98, 0.15, 0.33, 0.48)$. We remark that the numbers for $E$ are obtained here by  comparing the  full resolution original map with the half resolution reconstruction transformed to full resolution (replacing individual pixels by single-colored $2\times2\times2$ voxel blocks).

\subsection{3D tessellations}
Instead of solving multiple linear programs, one for each slice (possibly in parallel), we now solve for $d=3$ a single linear program that involves typically far more variables and constraints.  

For demonstration purposes we reconstruct with GBPD a \SI{240 x 240 x 35}{\micro\metre} subvolume of Data Set~I at half resolution (\SI{1.4}{\micro\metre} voxel size). This subvolume contains $109$ grains. The resulting linear program has \num[group-separator = {,}]{77828725} variables and \num[group-separator = {,}]{78542968} constraints (including the non-negativity constraints); a solution is found in around \SI{6}{\hour} of computation time on a similar machine but with $70$ GByte of RAM.

\begin{figure}[htb]
\subfigure[]{\includegraphics[height=0.3\textheight]{slice50gpd.pdf}}
\subfigure[]{\includegraphics[height=0.3\textheight]{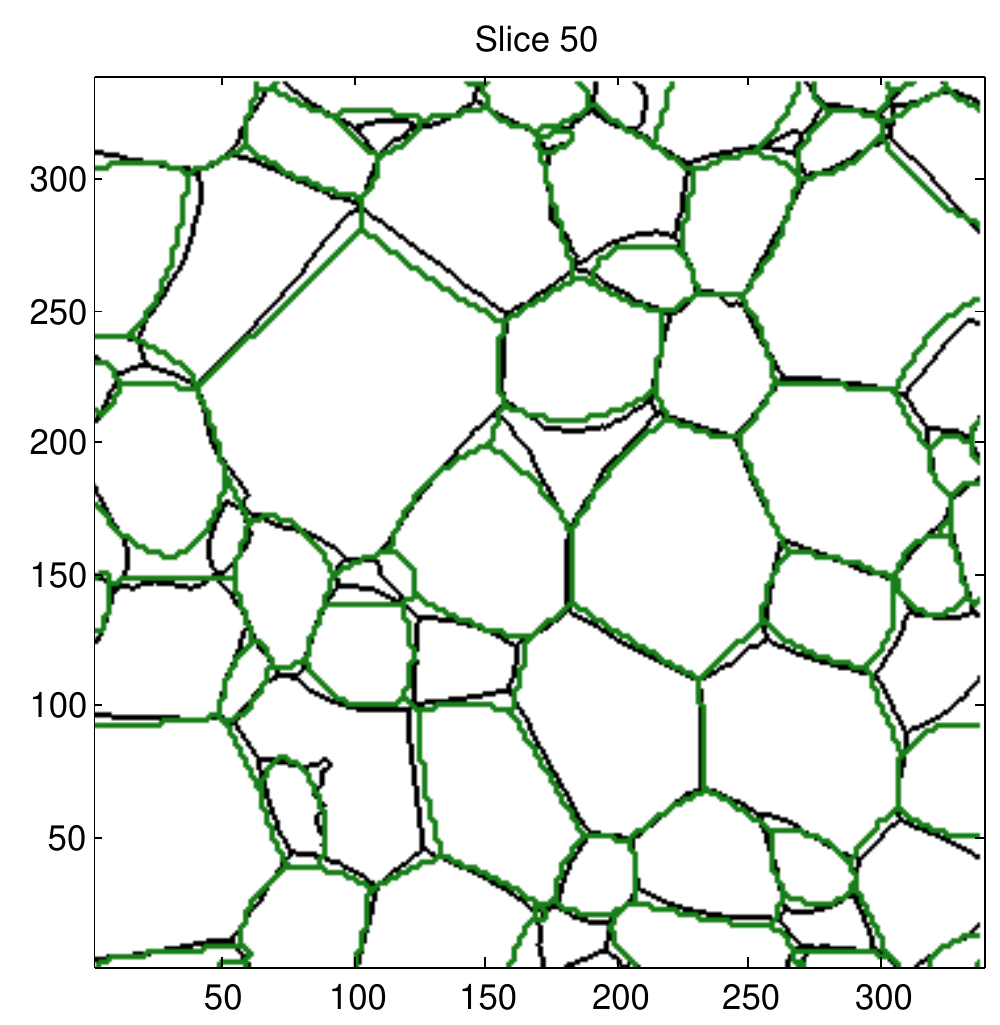}}
\caption{Generalized balanced power diagrams for slice 50 obtained at full resolution. (a) Reconstruction in 2D depicted in blue color, (b) reconstruction in 3D depicted in green color. The real grain boundaries are shown in black.}\label{fig9}
\end{figure}

We find that $93.78$ percent of the voxels are correctly labeled and that the number of erroneously reconstructed neighbors per grain is $0.70$ (in fact, $E=(93.78, 46.79, 84.40, 0.17, 0.53, 0.70)$). Fig.~\ref{fig9} shows the reconstruction for slice~50.

The GBPD approach for Data Set~I in 3D is seen to be inferior to its slice-by-slice counterpart. This
reflects in part the much fewer input parameters used in the 3D case: for a grain which is present in $100$ layers, in the 2D case we provide $600$ input parameters (CMS $(x,y)$, area, moments for each layer) while in the 3D case we provide only $10$ parameters (CMS $(x,y,z)$, volume, six moments). In addition it may partly be explained by the fact that the assumption of 3D convexity is stronger than slice-by-slice 2D convexity. 

\section{Discussion}
It is remarkable that the GBPD provides such excellent tessellations on our data sets. The physics underlying the two data sets (coarsening of Ti and plastic deformation and dynamical recovery of Al) is quite different and yet in both cases the parameterization seems adequate to obtain grains maps of a quality that is useful for a range of simulation tasks. Actually, with the layer-by-layer type of characterization---typical of 3DXRD---the quality is even sufficient to be used for statistics of grain topologies. Hence our experimental findings suggest that the physical principles that govern the forming of grain structures may comprise only a small number of degrees of freedom. Clearly, the long-term goal is to better understand this forming process.

The quality of the tessellations is also remarkable in view of the fact that the input sites $\vect{s_i}$ in~\eqref{eq:lp} are in close correspondence to the CMS of the reconstructed cells (see Table~\ref{table2}). Power diagrams with this property are called \emph{centroidal} \cite{briedengritzmann12}. The physical processes behind the formation of our data sets seem to favor the generation of such centroidal power diagrams. For more irregular grains, the relative distance between sites and CMS points may be larger. In such cases, we propose to globally optimize the sizes according to \cite{briedengritzmann10} or to run the LP algorithm in a loop, where new site positions are defined iteratively from the old CMS points as in \cite[Alg.~2]{bbg13} for a local optimization; this is also similar to the iterative step in the heuristic from \cite{laguerre11}. 

The linear programming formalism provides an optimal solution, something that is not guaranteed by heuristics.  However, the processing time is not negligible and the development of faster approximate solution methods may be relevant. 

In outlook, the GBPD approach can be generalized in several ways, including: 

\begin{itemize}
\item more complex shapes: the ellipsoidal norms $||\cdot||_A$ can be replaced by other gauge functions, representing more complex shapes that may incorporate a priori knowledge or other measured information.  This results only in a different objective function of the linear program that has to be solved. In fact, the parameters $\gamma_{i,j}$ in (LP) have to be adjusted.
\item weights: While we only used unit weights in our computations, the parameters $\omega_j$ in~\eqref{eq:lp} may in principle take arbitrary values. This can be used to speed up the algorithm by basing the computations on (virtual) adaptive resolutions. 
\end{itemize}

\section{Conclusions}\label{sect:conclusions}
By introducing the concept of generalized balanced power diagrams, we have demonstrated that a large class of grain structures can be represented remarkably well by tessellations that incorporate center of mass, volume (resp. areas), and second-order moments of the grains. An exact yet computationally feasible method for generating such tessellations was presented.

\section*{Acknowledgements}
We are grateful to an anonymous referee for pointing out the reference \cite{aljfs-14}. We further thank Oleg Mishin and Yubin Zhang for providing Data Set~II, and Erik Mejdal Lauridsen and Dorte Juul Jensen for valuable discussions. 

The Deutsche Forschungsgemeinschaft (DFG) is acknowledged for partial financial support through grant GR~993/10-2. HFP acknowledges the ERC advanced grant ``diffraction based transmission x-ray microscopy.'' COST Action MP1207 is acknowledged for networking support.


\begin{thebibliography}{10}

\bibitem{aljfs-14}
H.~Altendorf, F.~Latourte, D.~Jeulin, M.~Faessel, and L.~Saintoyant.
\newblock {3D} reconstruction of a multiscale microstructure by anisotropic
  tessellation models.
\newblock {\em Image Anal. Stereol.}, 33(2):121--130, 2014.

\bibitem{akl-13}
F.~Aurenhammer, R.~Klein, and D.-T. Lee.
\newblock {\em Voronoi diagrams and {D}elaunay triangulations}.
\newblock World Scientific, Singapore, 2013.

\bibitem{bleuet08}
P.~Bleuet, E.~Welcomme, E.~Dooryhee, J.~Susini, H.L. Hodeau, and P.~Walter.
\newblock Probing the structure of heterogeneous diluted materials by
  diffraction tomography.
\newblock {\em Nat. Mater.}, 7:468--472, 2008.

\bibitem{bwy-08}
J.-D. Boissonnat, C.~Wormser, and M.~Yvinec.
\newblock Anisotropic diagrams: the {L}abelle {S}hewchuk approach revisited.
\newblock {\em Theor. Comput. Sci.}, 408(2-3):163--173, 2008.

\bibitem{bbg13}
S.~Borgwardt, A.~Brieden, and P.~Gritzmann.
\newblock A balanced {$k$}-means algorithm for weighted point sets.
\newblock Submitted, {\tt arXiv:1308.4004 [math.OC]}, 2013.

\bibitem{briedengritzmann10}
A.~Brieden and P.~Gritzmann.
\newblock On clustering bodies: geometry and polyhedral approximation.
\newblock {\em Discrete Comput. Geom.}, 44(1):508--534, 2010.

\bibitem{briedengritzmann12}
A.~Brieden and P.~Gritzmann.
\newblock On optimal weighted balanced clusterings: gravity bodies and power
  diagrams.
\newblock {\em SIAM J. Discrete Math.}, 26(2):415--434, 2012.

\bibitem{validationCT}
P.~Cloetens, M.~Pateyron-Salom{\'e}, J.Y. Buffi{\`e}re, G.~Peix, J.~Baruchel,
  F.~Peyrin, and M.~Schlenker.
\newblock Observation of microstructure and damage in materials by phase
  sensitive radiography and tomography.
\newblock {\em J. Appl. Phys.}, 81(9):5878--5886, 1997.

\bibitem{dehoff85}
R.T. DeHoff and G.Q. Liu.
\newblock Grain topology.
\newblock {\em Metall. Trans. A}, 16:2007--2011, 1985.

\bibitem{duan2014}
Q.~Duan, D.P. Kroese, T.~Brereton, A.~Spettl, and V.~Schmidt.
\newblock Inverting {L}aguerre tessellations.
\newblock {\em Comput. J.}, 57(9):1431--1440, 2014.

\bibitem{Xpress}
{FICO}{\texttrademark}.
\newblock {Xpress Optimization Suite}.
\newblock \url{http://www.fico.com}.

\bibitem{dream3d2}
M.~Groeber, S.~Ghosh, M.D. Uchic, and D.M. Dimiduk.
\newblock A framework for automated analysis and simulation of {3D}
  polycrystalline microstructures. {P}art 2: {S}ynthetic structure generation.
\newblock {\em Acta Mater.}, 56(6):1274--1287, 2008.

\bibitem{dream3d1}
M.A. Groeber and M.A. Jackson.
\newblock {DREAM.3D: A} digital representation environment for the analysis of
  microstructure in {3D}.
\newblock {\em Integr. Mater. Manuf. Innov.}, 3(1):5, 2014.

\bibitem{hefferan09}
M.~Hefferan, S.F. Li, J.~Lind, U.~Lienert, A.D. Rollett, P.~Wynblatt, and R.M.
  Suter.
\newblock Statistics of high purity nickel microstructure from high energy
  x-ray diffraction microscopy.
\newblock {\em CMC-Comput. Mater. Con.}, 14:209--219, 2009.

\bibitem{ice11}
G.E. Ice, J.D. Budai, and J.W. Pang.
\newblock The race to x-ray microbeam and nanobeam science.
\newblock {\em Science}, 334:1234--1239, 2011.

\bibitem{king08}
A.~King, G.~Johnson, D.~Engelberg, W.~Ludwig, and J.~Marrow.
\newblock Observations of intergranular stress corrosion cracking in a
  grain-mapped polycrystal.
\newblock {\em Science}, 321:382--385, 2008.

\bibitem{kuehn08}
M.~K\"uhn and M.O. Steinhauser.
\newblock Modeling and simulation of microstructures using power diagrams:
  {P}roof of the concept.
\newblock {\em Appl. Phys. Lett.}, 93(3):034102--034102--3, 2008.

\bibitem{ls-03}
F.~Labelle and J.R. Shewchuk.
\newblock Anisotropic {V}oronoi diagrams and guaranteed-quality anisotropic
  mesh generation.
\newblock In {\em SCG '03: Proceedings of the nineteenth annual symposium on
  Computational geometry}, pages 191--200. ACM Press, 2003.

\bibitem{larson02}
B.C. Larson, W.~Yang, G.E. Ice, J.D. Budai, and T.Z. Tischler.
\newblock Three-dimensional x-ray structural microscopy with submicrometre
  resolution.
\newblock {\em Nature}, 415:887--890, 2002.

\bibitem{levine06}
L.E. Levine, B.C. Larson, W.~Yang, M.E. Kassner, J.Z. Tischler, M.A.
  Delos-Reyes, R.J. Fields, and W.~Liu.
\newblock X-ray microbeam measurements of individual dislocation cell elastic
  strains in deformed single-crystal copper.
\newblock {\em Nat. Mater.}, 5:619--622, 2006.

\bibitem{liu11}
H.H. Liu, S.~Schmidt, H.F. Poulsen, A.~Godfrey, Z.Q. Liu, J.A. Sharon, and
  X.~Huang.
\newblock Three-dimensional orientation mapping in the transmission electron
  microscope.
\newblock {\em Science}, 332:833--834, 2011.

\bibitem{ludwig08}
W.~Ludwig, S.~Schmidt, E.M. Lauridsen, and H.F. Poulsen.
\newblock X-ray diffraction contrast tomography: a novel technique for
  three-dimensional grain mapping of polycrystals. {I.} direct beam case.
\newblock {\em J. Appl. Cryst.}, 41:302--309, 2008.

\bibitem{phdallan}
A.~Lyckegaard.
\newblock {\em Development of tomographic reconstruction methods in materials
  science with focus on advanced scanning methods}.
\newblock PhD thesis, Ris{\o}¸ National Laboratory for Sustainable Energy,
  Technical University of Denmark, 2011.

\bibitem{lyckegard10}
A.~Lyckegaard, A.~Alpers, W.~Ludwig, R.W. Fonda, L.~Margulies, A.~G{\"o}tz,
  H.O. S{\o}rensen, S.R. Deyk, H.F. Poulsen, and E.M. Lauridsen.
\newblock {3D} grain reconstruction from boxscan data.
\newblock In {\em Proceedings of the 31st Ris{\o} International Symposium on
  Materials Science: Challenges in materials science and possibilities in 3D
  and 4D characterization techniques}, pages 329--336, 2010.

\bibitem{laguerre11}
A.~Lyckegaard, E.M. Lauridsen, W.~Ludwig, R.W. Fonda, and H.F. Poulsen.
\newblock On the use of {L}aguerre tessellations for representations of {3D}
  grain structures.
\newblock {\em Adv. Eng. Mater.}, 13(3):165--170, 2011.

\bibitem{mckenna14}
I.M. McKenna, S.O. Poulsen, E.M. Lauridsen, W.~Ludwig, and P.W. Voorhees.
\newblock Grain growth in four dimensions: A comparison between simulation and
  experiment.
\newblock {\em Acta Mater.}, 78:125--134, 2014.

\bibitem{mikadawson98}
P.M. Mika and P.R. Dawson.
\newblock Effects of grain interaction on deformation in polycrystals.
\newblock {\em Mater. Sci. Eng.}, A257(1):62--76, 1998.

\bibitem{oleg2012}
O.V. Mishin, V.M. Segal, and S.~Ferrasse.
\newblock Quantitative microstructural characterization of thick aluminum
  plates heavily deformed using equal channel angular extrusion.
\newblock {\em Metall. Mater. Trans. A}, 43:4767--4776, 2012.

\bibitem{additional1}
J.~Oddershede, S.~Schmidt, H.F. Poulsen, L.~Margulies, J.~Wright, M.~Moscicki,
  W.~Reimers, and G.~Winther.
\newblock Grain-resolved elastic strains in deformed copper measured by
  three-dimensional x-ray diffraction.
\newblock {\em Mater. Charact.}, 62:651--660, 2011.

\bibitem{papadimitriou98}
C.H. Papadimitriou and K.~Steiglitz.
\newblock {\em Combinatorial optimization: algorithms and complexity}.
\newblock Dover Publications, Inc., Mineola, 1998.

\bibitem{patterson13}
B.R. Patterson, D.J. Rule, R.T. DeHoff, and V.~Tikare.
\newblock Schlegel description of grain form evolution in grain growth.
\newblock {\em Acta Mater.}, 61, 2013.

\bibitem{poulsennielsen01}
H.F. Poulsen, S.F. Nielsen, E.M. Lauridsen, S.~Schmidt, R.M. Suter, U.~Lienert,
  L.~Margulies, and T.~Lorentzen D.~Juul Jensen.
\newblock Three-dimensional maps of grain boundaries and the stress state of
  individual grains in polycrystals and powders.
\newblock {\em J. Appl. Cryst.}, 34:751--756, 2001.

\bibitem{additional2}
P.~Reischig, A.~King, L.~Nervo, N.~Vigan{\'o}, Y.~Guilhem, W.J. Palenstijn,
  K.J. Batenburg, M.~Preuss, and W.~Ludwig.
\newblock Advances in x-ray diffraction contrast tomography: flexibility in the
  setup geometry and application to multiphase materials.
\newblock {\em J. Appl. Cryst.}, 46:297--311, 2013.

\bibitem{riosglicksman07}
P.R. Rios and M.E. Glicksman.
\newblock Topological and metrical analysis of normal grain growth in three
  dimensions.
\newblock {\em Acta Mater.}, 55(1):1565--1571, 2007.

\bibitem{laguerre2}
A.~Spettl, T.~Werz, C.E.~Krill III, and V.~Schmidt.
\newblock Parametric representation of {3D} grain ensembles in polycrystalline
  microstructures.
\newblock {\em J. Stat. Phys.}, 154:913--928, 2014.

\bibitem{telley96a}
H.~Telley, T.M. Liebling, and A.~Mocellin.
\newblock The {L}aguerre model of grain growth in two dimensions: {I. C}ellular
  structures viewed as dynamical {L}aguerre tessellations.
\newblock {\em Philos. Mag. B}, 73(3):395--408, 1996.

\bibitem{telley96b}
H.~Telley, T.M. Liebling, and A.~Mocellin.
\newblock The {L}aguerre model of grain growth in two dimensions: {II.
  E}xamples of coarsening simulations.
\newblock {\em Philos. Mag. B}, 73(3):409--427, 1996.

\bibitem{telley92}
H.~Telley, T.M. Liebling, A.~Mocellin, and F.~Righetti.
\newblock Simulating and modelling grain growth as the motion of a weighted
  {V}oronoi diagram.
\newblock {\em Mater. Sci. Forum}, 94-96(1):301--306, 1992.

\bibitem{thortonpoulsen08}
K.~Thorton and H.F. Poulsen.
\newblock Three-dimensional materials science: An intersection of
  three-dimensional reconstructions and simulations.
\newblock {\em MRS Bull.}, 33:587--595, 2008.

\bibitem{uchic06}
M.D. Uchic, M.A. Groeber, D.M. Dimiduk, and J.P. Simmons.
\newblock {3D} microstructural characterization of nickel superalloys via
  serial-sectioning using a dual beam {FIB-SEM}.
\newblock {\em Scripta Mater.}, 55:23--28, 2006.

\bibitem{xue97}
X.~Xue, F.~Righetti, H.~Telley, and T.M. Liebling.
\newblock The {L}aguerre model for grain growth in three dimensions.
\newblock {\em Philos. Mag. B}, 75(4):567--585, 1997.

\bibitem{zaefferer08}
S.~Zaefferer, S.I. Wright, and D.~Raabe.
\newblock {3D}-orientation microscopy in a {FIB SEM}: A new dimension of
  microstructural characterization.
\newblock {\em Metall. Mater. Trans. A}, 39:374--389, 2008.

\end{thebibliography}

\end{document}